\documentclass[a4paper,11pt]{article}
\usepackage{pos}

\title{Recent progress on nucleon form factors}

\author*[a]{Dalibor Djukanovic}

\affiliation[a]{Helmholtz Institute Mainz, Staudingerweg 18, D-55128 Mainz,
Germany\\
GSI Helmholtzzentrum f\"ur Schwerionenforschung, D-64291 Darmstadt, Germany}

  \emailAdd{d.djukanovic@him.uni-mainz.de}

\abstract{The form factors of the nucleon provide key information on nucleon
properties. When confronted with precisely measured observables from
experiments, they serve as benchmark quantities for lattice calculations. On the
other hand lattice determinations may serve as vital theory input for the
interpretation of experiments, e.g. in neutrino-nucleus scattering. I 
review recent progress in the calculation of nucleon form factors on the
lattice and its relevance to future experiments.}

\FullConference{%
 The 38th International Symposium on Lattice Field Theory, LATTICE2021
  26th-30th July, 2021
  Zoom/Gather@Massachusetts Institute of Technology
}

\begin{document}
\maketitle

\section{Introduction}
Form factors encode fundamental properties of the nucleon, parametrizing  its
response to external currents, which are uniquely defined through their quantum numbers. 
We can derive basic
properties of the nucleon from these form factors, e.g. charges, charge
distributions, stiffness or rigidity. Thus a first principle calculation from
lattice Quantum Chromodynamics (QCD) gives us a key insight into the strong
interactions governing the forces inside nucleons. A detailed knowledge of the
nucleon form factor is vital to the success of upcoming high precision experiments
involving nuclear targets like DUNE at Fermilab \cite{DUNE:2015lol} or
Hyper-Kamiokande \cite{Hyper-Kamiokande:2018ofw}. Recent progress in
nucleon form factor calculations suggests a number of areas, where the lattice may
have an immediate impact, e.g.
\begin{itemize}
	\item in searches for beyond Standard Model (BSM) physics,
	\item as high precision input in analysis of  experimental
		data,
	\item in cases where there is disagreement between different
		experiments.
\end{itemize}
The first point is mostly the domain of the nucleon form factors at vanishing
momentum transfer (charges). Restricting models of
BSM physics one needs the corresponding matrix elements of non-Standard Model
hadronic currents in order to  
establish strong bounds from experiments (c.f.
\cite{Bhattacharya:2011qm}). Moreover, the axial charge is very well determined
experimentally and serves as a benchmark quantity for extraction techniques on the
lattice, where remarkable precision has been reached \cite{Chang:2018uxx}. In dark matter searches the
matrix element of the scalar current, the sigma term, plays an important role.
There is a particularly interesting connection between the sigma term and $\pi$N
scattering, via the Cheng-Dashen theorem \cite{Cheng:1970mx}, that enables
direct comparisons of this quantity to dispersive (experimental data driven)
determinations \cite{Hoferichter:2015dsa}. The sigma term is especially
intriguing, since there is a slight tension between the dispersive analysis of
\cite{Hoferichter:2015dsa} and the $N_F=2+1$ average of lattice determinations
\cite{Aoki:2021kgd}. A very recent analysis
\cite{Gupta:2021ahb} suggests that the tension might be due to excited-state
contributions highlighting not only the need for statistical precision but for a
high level of control over systematics. 

Also the
case of non-vanishing momentum transfer is highly interesting, not only for 
nucleon properties, but also in BSM physics searches and high precision determinations
of SM observables at low energies. In the upcoming experiments at DUNE 
\cite{DUNE:2015lol} the axial form factor of the nucleon plays a crucial role for the
interpretation of the data, especially in the region, where quasi-elastic
neutrino-nucleus scattering is the dominant process (c.f.
\cite{Kronfeld:2019nfb}). A determination of the axial radius to 20\% accuracy
is sufficient to render the theoretical uncertainty due to $r_A$ in neutron
quasi-elastic cross sections to a subdominant contribution \cite{Hill:2017wgb}.
Here not only the radius is of interest but rather the whole $Q^2$ dependence of
the form factor.

The increase in 
precision opens up new windows of opportunity to determine  SM parameters, such as the Weinberg
angle, in low energy experiments, e.g. by
measuring the weak charge of the proton in parity-violation
experiments like  P2 at MESA \cite{hug:erl2019-mocoxbs05} or Q-weak at JLAB.
Here a detailed knowledge of the strange electromagnetic form factors is an important
ingredient in the extraction of the weak charge \cite{Becker:2018ggl}. 

One of the opportunities where lattice determinations may hope to resolve a
persisting discrepancy is the proton radius. For the proton radius experimental data from
$ep$-scattering \cite{A1:2010nsl} and spectroscopy measurements of muonic hydrogen Refs.
\cite{Pohl:2010zza,Antognini:2013txn} are at odds. A recent measurement of the proton
radius, again from $ep$-scattering, seems to favor the smaller radius 
\cite{Xiong:2019umf}, which is also consistent with dispersive
analysis (c.f \cite{Lin:2021umz}). 

The impact of the lattice crucially depends on the achievable accuracy, not only
in terms of statistical precision, but to which degree all relevant systematics are
understood and under control. There has been quite some progress in the last few
years concerning both and I review the current status of affairs, focusing
on the vector- and axialvector form factors. 

In an effort to make this proceeding self-contained I first give an overview of the 
methods used in the extraction of the nucleon form factors. For a
very recent more detailed account I refer to last years' conference proceeding
\cite{Ottnad:2020qbw}.  In the third section I highlight possible sources of
systematic uncertainty. Finally I summarize the most recent result
for the vector and axialvector form factors.

\section{Lattice methodology}
The quantities calculated on the lattice are euclidean $n$-point correlation
functions of hadronic operators, where the nucleon is created (annihilated) via an interpolating
source (sink) operator
$\overline{\Psi} (\Psi)$ typically of the form
\begin{align}
\Psi_\alpha(x) = \epsilon_{abc}
\left({u}^T_a(x)C\gamma_5{d}_b(x)\right){u}_{c,\alpha}(x)\,,
	\label{nucleon_interpolating_operator}
\end{align}
where $C$ is the charge conjugation matrix, $u$ and $d$ denote the up and down
quarks. 

In the time-momentum representation 
the spin-projected two-point function is then given by
\begin{align}
	C_2(t;{\bf{p}})&=\Gamma_{\alpha\beta} \sum\limits_{{\bf{x}}} e^{-i{\bf{px}}}\Bigl\langle\Psi_\beta({\bf{x}},t) \overline{\Psi}_\alpha (0) \Bigr\rangle,
\end{align}
where the source is shifted to the origin. Inserting a complete set of
energy-eigenstates the spectral decomposition of the two-point function reads
\begin{align}
	C_2(t;{\bf{p}})=\sum\limits_n  \underbrace{|\langle n |\Psi| N
	\rangle|}_{Z_n}
	e^{-E_n t},
\end{align}
where all states $|n\rangle$  compatible with the quantum numbers 
of the interpolating operator, i.e. beyond the ground state also
excited and multi-particle states, contribute. The exponential falloff leads to a
suppression of excited states for large enough distances between source and
sink. In this region the
dominant contribution to the correlation functions comes from the ground state
and its properties are readily
read off, e.g. as plateaus in effective mass plots defined via
\begin{align}
	E_{\mathrm{eff}}&= \frac{1}{\tau}\ln
	\frac{C_2(t;{\bf{p}})}{C_2(t+\tau;{\bf{p}})}.\
	\label{effective_mass}
\end{align}

Unfortunately baryons are affected by an infamously unfavorable signal to noise
ratio, which effectively drowns the signal in noise once the ground state would
start to dominate. One may intuitively understand this looking at the variance
directly calculated using the interpolating operator of
Eq.~(\ref{nucleon_interpolating_operator}) (c.f.  Ref.~\cite{Lepage:1989hd}).
One possible contraction of the squared operator results in three pions
propagating from source to sink, thus the variance in the large time separation
limit reads
\begin{align}
	\frac{\Delta C_2(t;{\bf{0}})}{C_2(t;{\bf{0}})} \sim
	\frac{\exp(-\frac{3}{2} m_\pi
	t)}{\exp (-m_n t)} .
	\label{signal_to_noise}
\end{align}
Therefor in the asymptotic limit the noise grows exponentially. In order to reach the
ground state region earlier, effectively taming the signal-to-noise problem, smearing
techniques are used which lead to an increased overlap of the interpolating
operator with the ground state
\cite{Gusken:1989qx,APE:1987ehd,vonHippel:2013yfa}. 
The signal-to-noise problem is exacerbated in the case of the
three-point functions, where the computationally feasible source sink separations are severely
limited by the dramatic increase in cost. Moreover in the contractions pertinent
to form factor calculations, so called
quark-disconnected contributions may arise, which are notoriously difficult to
calculate. Most simulations use degenerate light quarks, which in some  of the 
isovector combinations happen to be free of
quark-disconnected contributions. However this is not true for all, e.g. the
strange electromagnetic form factor, and for the flavor decomposition the isoscalar
form factors are still needed, which receive contributions from these types of
diagrams.  For the calculation of the connected diagrams most analysis use the
sequential inversion method with a fixed sink, i.e. located at a fixed time
separation to the source. While in this setup for  every source-sink separation and every sink
momentum an explicit inversion is needed, the correlator for all operator
insertions between source and sink are accessible without further inversion. The
statistical precision is usually increased using variance
reduction technique such as all mode averaging \cite{Bali:2009hu,Shintani:2014vja}.
In recent years algorithmic developments, e.g. hierarchical probing
\cite{Stathopoulos:2013aci}, low mode deflation \cite{Gambhir:2016uwp},
frequency-splitting \cite{Giusti:2019kff} to name few, have cut the cost for the
calculation of disconnected diagrams dramatically.

For the form factor calculations  the extractions usually proceed via the ratio of
two- and three-point functions, where the ratio is constructed such that overlap
factors cancel and the ground state matrix element is dominant.
\begin{figure}[t]
	\includegraphics[width=.95\textwidth]{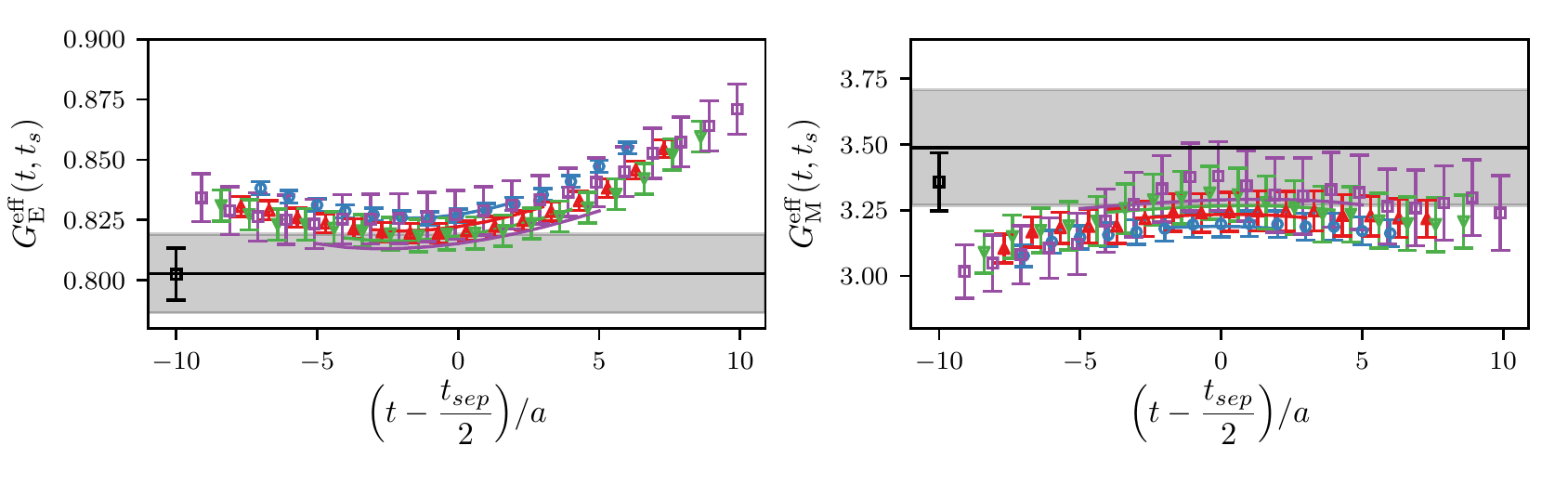}
	\caption{Effective form factor for the electric (left) and magnetic
		(right) isovector vector current, on ensemble D200 at first
	non-vanishing momentum transfer (figure taken from
Ref.~\cite{Djukanovic:2021cgp}). The effective form factor is plotted for
different values of $t_{\text{sep}}$ between 1 and 1.5 fm, together with the
estimate of ground state matrix element from the direct fits including excited states
(black data points) and the summation method result (gray band).}
\label{fig_eff_d200}
\end{figure}

In most studies one of the following 
ratios is used \cite{Draper:1989pi,Alexandrou:2008rp}  
\begin{align}
R^{X}(t,t_s;{\bf{q}})&=\frac{C_3^{X}(t,t_s;{\bf{q}})}{C_2(t_s;{\bf{0}})} 
\sqrt{\frac{C_2(t_s-t;-{\bf{q}})\, C_2(t,{\bf{0}})\, C_2(t_s;{\bf{0}})}{C_2(t_s-t;{\bf{0}})\, 
C_2(t;-{\bf{q}}) \,C_2(t_s;-{\bf{q}})} } \, , \label{ratio1}\\
R^{X}(t,t_s;{\bf{q}})&=\frac{C_3^{X}(t,t_s;{\bf{q}})}{C_2(t_s;{\bf{0}})} ,
\label{ratio2}
\end{align}
where the nucleon three-point-function of a general operator $X$ is  given by
\begin{align}
C_3^{X}(t,t_s;{\bf{q}})&=\Gamma_{\alpha\beta} \sum\limits_{{\bf{x,y}}} e^{i{\bf{qy}}}
\Bigl\langle\Psi_\beta({\bf{x}},t_s) X({\bf{y}},t) \overline{\Psi}_\alpha (0)
\Bigr\rangle.
\end{align}
For vanishing momentum transfer the ratios coincide, however for non-vanishing
momenta the overlap factors of the two-point functions do not cancel for the
latter. Matching the spectral representation of the ratios in
Eqs.~(\ref{ratio1},\ref{ratio2}) to the corresponding nucleon matrix elements
parameterized using form factors, one obtains effective form factors (see
Fig.~\ref{fig_eff_d200}).

In the fixed sink method the nucleon at the sink is usually at rest,
i.e. for a momentum transfer ${\bf{q}}$ the initial and final nucleon states have momenta
\begin{align}
{\bf{p'}}=0,\qquad {\bf{p}}=-{\bf{q}}. 
\end{align}
The
operator $X$ is the current operator, which is classified with respect to
its symmetry
as 
\begin{align}
	X^V_\mu(x)&=\bar q(x) \gamma_\mu q(x),\\
	X^A_\mu(x)&= \bar q(x) \gamma_\mu \gamma_5 q(x),\\
	X^S(x)&= \bar q(x) q(x),\\
	X^T_{\mu\nu}(x)&= \bar q(x) \sigma_{\mu\nu} q(x) ,
	\label{operator}
\end{align}
for the vector, axial, scalar and tensor currents, respectively. Also non-local
operators are used, e.g. point-split currents in the vector case resulting in conserved charges.
\section{Sources of systematics}
\begin{figure}[t]
	\begin{center}
		\includegraphics[width=.45\textwidth]{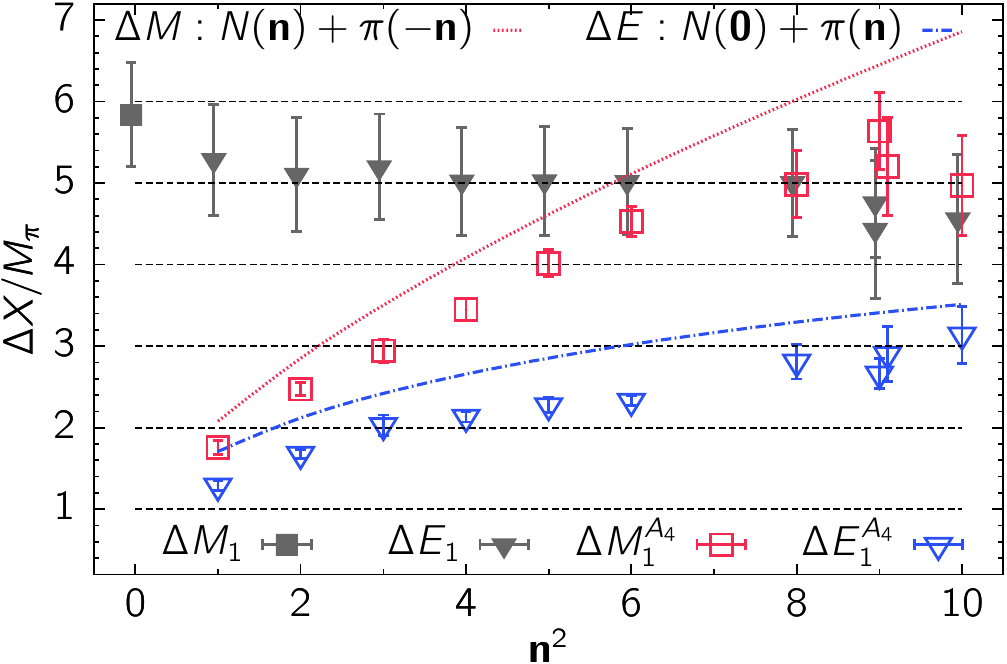}
		\includegraphics[width=.45\textwidth]{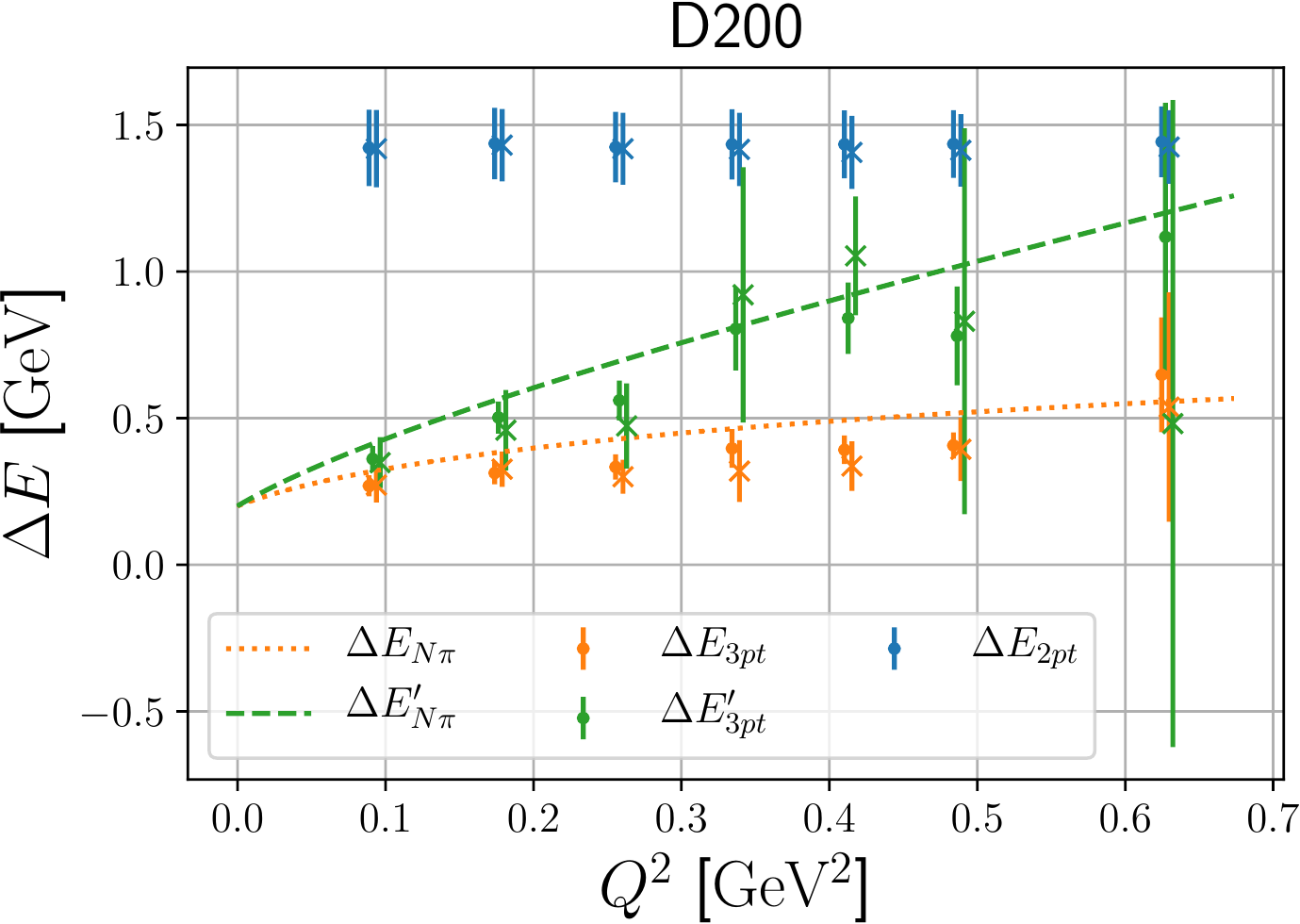}
	\end{center}
	\caption{Plots showing the different values of the energy gap to the
		first excited state, when fitted to the two- or three-point
		functions. The figures originally were published in \cite{Jang:2019vkm} (left) and
		\cite{RQCD:2019jai} (right), and are reproduced under the  Creative Commons
	Attribution 4.0 International license.}

	\label{fig_diff_egap}
\end{figure}

The most severe problem every lattice calculation of the nucleon form factors
faces is that of excited states. While in the asymptotic limit of large
euclidean time separations ratios like Eq.(\ref{ratio1})  are proportional to
the ground state hadronic matrix element we are after, the computationally
affordable $t_{\mathrm{sep}}$ are such that there still is sizable contamination
left from excited states (see Fig.~\ref{fig_eff_d200}). One obvious remedy is to
simply keep more terms in the spectral decomposition of the two- and three-point
functions
\begin{align}
	C_2(t;{\bf{p}})&= |Z_0({\bf{p}})|^2 \exp \bigl[ - E_0({\bf{p}}) t \bigr] +
	|Z_1({\bf{p}})|^2 \exp\bigl[ -E_1({\bf{p}}) t \bigr]+ \ldots \, ,\\
	C_3(t,t_{\mathrm{sep}};{\bf{p'}},{\bf{p}}) &=    Z_0({\bf{p'}})
	Z_0^*({\bf{p}}) \langle 0 | X | 0 \rangle \exp\bigl[ -
	E_0({\bf{p'}})(t_{\mathrm{sep}} -t)\bigr] \exp\bigl[
		-E_0({\bf{p}}) t\bigr] \nonumber\\
		&+ Z_0({\bf{p'}}) Z_1^*({\bf{p}}) \langle 0 | X |1 \rangle \exp\bigl[ -
		E_0({\bf{p'}})(t_{\mathrm{sep}} -t)\bigr] \exp\bigl[
                -E_1({\bf{p}}) t\bigr] \nonumber\\
		&+ Z_1({\bf{p'}}) Z_0^*({\bf{p}}) \langle 1 | X |0 \rangle \exp\bigl[ -
		E_1({\bf{p'}})(t_{\mathrm{sep}} -t)\bigr] \exp\bigl[
                -E_0({\bf{p}}) t\bigr] \nonumber\\
		&+ Z_1({\bf{p'}}) Z_1^*({\bf{p}}) \langle 1 | X |1 \rangle \exp\bigl[ -
		E_1({\bf{p'}})(t_{\mathrm{sep}} -t)\bigr] \exp\bigl[
		-E_1({\bf{p}}) t\bigr]+\ldots \, ,
\end{align}
to fit all available data. In principle the overlaps $Z_i$ and the energies
$E_i$ are universal and one may try to fix these in fits to the two-point
functions first, in turn to be used as \emph{fixed parameters} for higher $n$-point
functions \cite{Jang:2019jkn}. Obtaining the excited-state
parameters from fits to the two-point functions is a daunting task given the
signal-to-noise problem. There is evidence that, at least for
some observables, this procedure might not be best suited to account for the
excited-state effects found in the three-point functions. The excited state
properties as extracted from the two- and three-point functions can differ
significantly,
 as has been observed especially in the axialvector sector
\cite{Jang:2019vkm,RQCD:2019jai,Alexandrou:2020okk} (see
Fig.~\ref{fig_diff_egap}). 
One alternative is to perform
simultaneous fits of two- and three-point functions fitting the excited
state properties as nuisance parameters. Given the large number of data points
that enter in such fits estimating the covariance matrix is very challenging and
the resulting procedure may become unstable. Also the current level of
statistics might not be sufficient to consistently extract all the excited-state
parameters from two- and three-point functions.
Alternatively an intermediate approach between fixing the excited state
parameters completely from the two-point functions and simultaneous fits leaving
them open is to use prior knowledge about the energies and overlaps,
with some freedom for the higher $n$-point function to still choose a different value.
While the gaps and overlaps may be universal it is
by no means clear that the correlation functions exhibit that pattern. We could
find ourselves in a situation where a multitude of excited states might be subsumed
into one \emph{effective} excited state that bares no resemblance to the one
from the two-point functions. The prior information might come from a simple
ansatz about the excitation spectrum of the interpolating operators, e.g.
non-interacting multi-particle states of nucleons and pions or refinements
including interactions \cite{Hansen:2016qoz}. One may also resort to estimates based on chiral
perturbation theory about the size of the excited state contaminations
\cite{Tiburzi:2015tta,Bar:2015zwa} which predicts the dominant contribution to come from pion-nucleon
states\footnote{In ratios like Eq.~(\ref{ratio1}) no new parameter enter the ChPT
	prediction of excited states if the smearing size of the interpolating
	operator is small compared to the inverse pion mass 
\cite{Bar:2015zwa}.}. In fact the energy of the first excited state as
extracted from the two-point functions
reported in recent studies is much higher than expected from a chiral analysis
and is closer to the Roper mass $N(1440)$, e.g. \cite{Alexandrou:2020okk}, or
even heavier resonances like 
$N(1710)$ \cite{Jang:2019vkm}.  Most recently the excited state analysis has
been extended to non-vanishing momentum transfer for the axialvector \cite{Bar:2018xyi} and
vector currents \cite{Bar:2021crj}, where the 
induced pseudoscalar form factor $G_P$ is shown to receive strong corrections.
While these analyses work best for larger source-sink separation than usually
 available, they still serve as a good indication about the expected size of these
corrections.

\begin{figure}[t]
	\includegraphics[width=.5\textwidth]{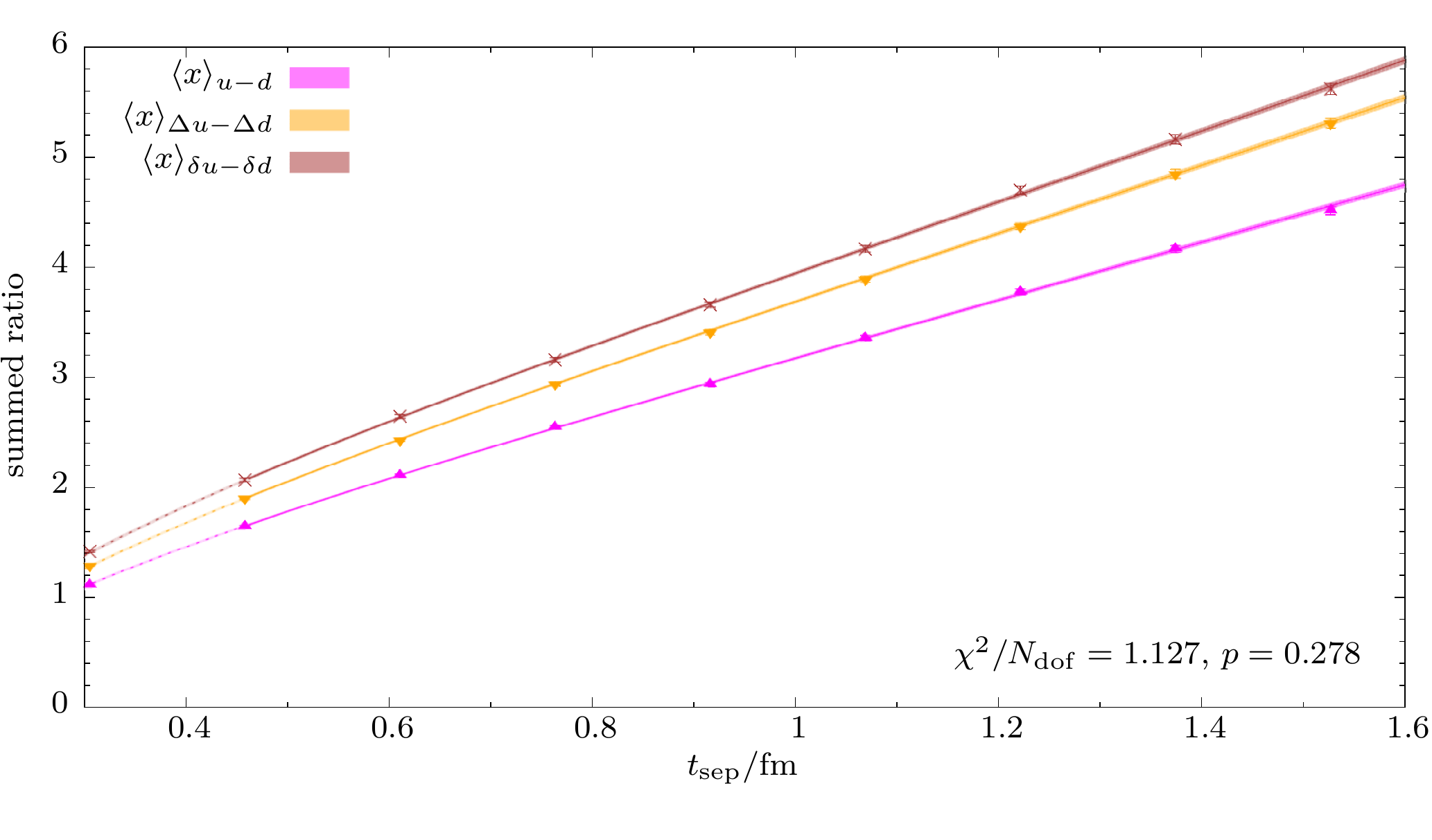}
	\includegraphics[width=.5\textwidth]{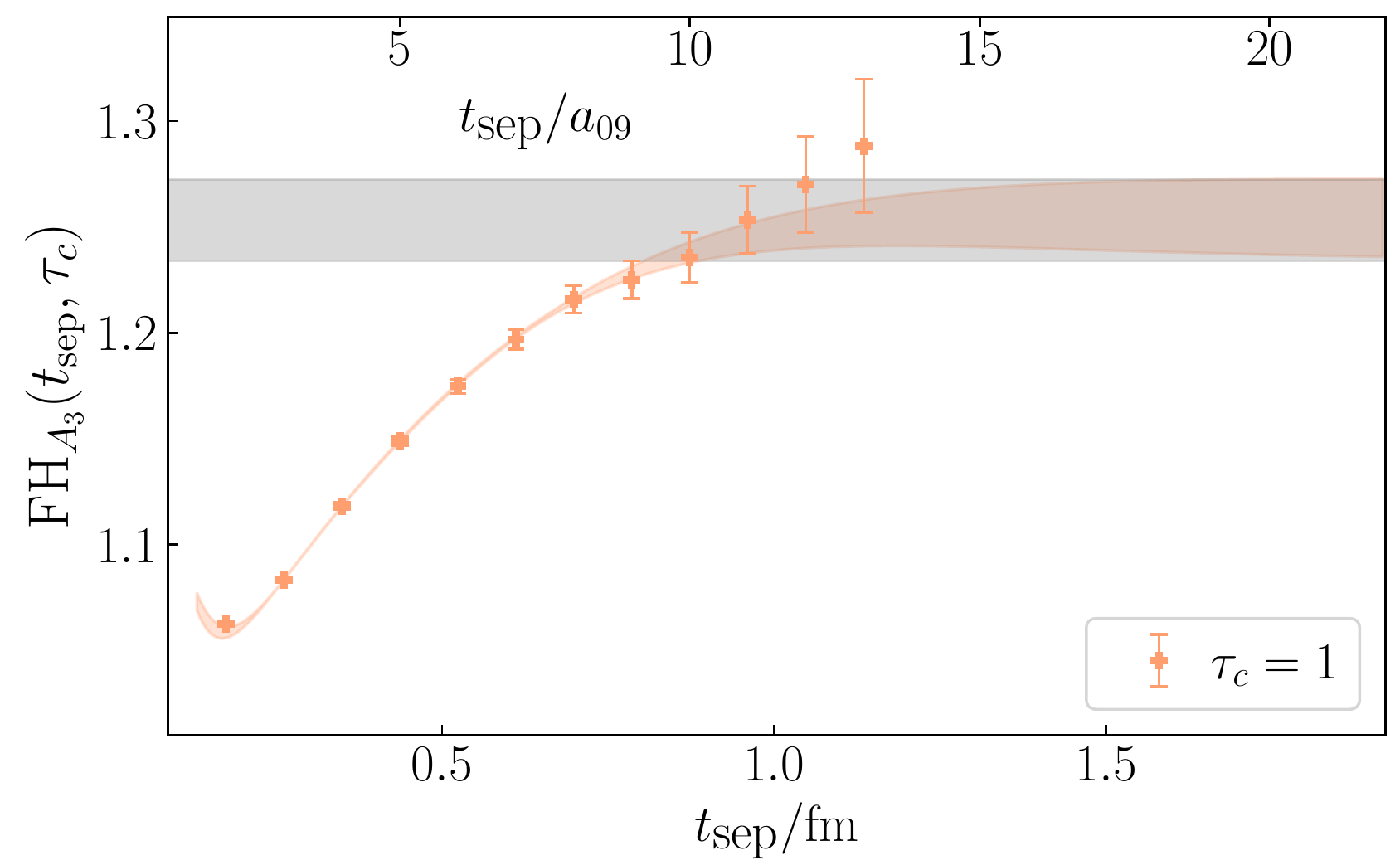}
	\caption{Examples for the extension of the summation method to smaller
		values of $t_\text{sep}$ for the twist-2 operators (left) and
		the Feynman-Hellmann correlator of the third component of the
		axial current \cite{He:2021yvm}. The figures originally were published in this conference
		proceedings \cite{Ottnad:2021tlx} (left) and
		in Ref.~\cite{He:2021yvm} (right), and are reproduced under the  Creative Commons
	Attribution 4.0 International license.}
	\label{fig_summation}
\end{figure}

Another approach dealing with excited states is given by the summation method
\cite{Maiani:1987by}, which has a long history in the calculation of nucleon
form factors \cite{Dong:1997xr,Capitani:2012gj,Green:2014xba,Capitani:2015sba,Capitani:2017qpc}.
Here one sums the correlators for timeslices between source and sink, which leads
to a parametric suppression of excited states with $t_\mathrm{sep}$ instead of
$t,(t_\mathrm{sep}-t)$. To illustrate this let us write a correlation function
that has been constructed such that the leading exponential falloff is cancelled
and $\Delta(\Delta')$ denote the energy gap of the excited  to the ground
state emanating from source (sink), i.e.
\begin{align}
	C(t,t_\mathrm{sep};{\bf{p}}) &= C_0 + C_1 \exp\bigl[- \Delta  t\bigr] +c
	\exp \bigl[- \Delta' (t_\mathrm{sep}-t) \bigr]
	\label{eq_groundstate_corr}\\
	S(t_\mathrm{sep}) &=
	\sum\limits_{t=t_{\mathrm{skip}}}^{t_{\mathrm{sep}}-t_{\mathrm{skip}}}
	C(t,t_\mathrm{sep};{\bf{p}})= C_0 \frac{t_\mathrm{sep}-2t_\mathrm{skip}
	+a }{a} \nonumber\\
	&+ C_1
	\frac{\exp\bigl[ -\Delta t_{\mathrm{skip}} \bigr] - \exp\bigl[ -\Delta
	(t_{\mathrm{sep}}-t_{\mathrm{skip}} +a ) \bigr] }{1- \exp\bigl[-\Delta
	a\bigr]} + \ldots \, ,\label{eq_summed_ratio}
\end{align}
where in the last line we see that the exponential falloff is enhanced compared
to Eq.~(\ref{eq_groundstate_corr}). One welcome simplification of this method is the
reduction in data size, leading to more stable estimates of covariance matrices.
Moreover, taking the derivative of Eq.~(\ref{eq_summed_ratio}) with respect to
$t_{\text{sep}}$ the region of
ground state dominance is marked by a plateau and one has the opportunity to
\emph{monitor} the window for which a stable extraction is possible even for
lower values of $t_\text{sep}$
\cite{He:2021yvm} (see Fig.~\ref{fig_summation}).  
In any case the identification of the ground state matrix element is a
challenging task. For some observables one may resort to constraints based on
symmetries, that various extractions of ground state matrix elements have to
fulfill once the asymptotic regime has been reached. One such constraint is given
by the PCAC relation, which has to be fulfilled on the correlator level. Any
deviation in such relations for the extracted ground state matrix elements may
be indicative for a failure to cleanly separate the excited-state contributions
(c.f. \cite{Jang:2019vkm,RQCD:2019jai,Park:2021ypf}). For observables where
symmetry constraints are not easily established the best option is to look
for a consistent value through various methods. 

Besides the excited-state contributions one major source of uncertainty is due
to discretization and  the finite size of the simulated boxes. A controlled
extrapolation to the continuum and infinite volume has to be performed. 
Another immediate consequence of the discretization is that momenta are
not continuous. When we are interested in the $Q^2$ dependence of the form
factors, or quantities defined via the slope (at vanishing momentum), an
interpolation of the simulated points is needed. This is also true in the
continuum when one has to fit the available experimental data. It is natural to
use the same approaches
as in experiments, and indeed most analysis adopted multiple strategies
ranging from historically motivated  fit
forms like dipole, or Pad\'e fits, model-independent approaches like
z-expansion \cite{Hill:2010yb} or effective field theory approaches using chiral
perturbation theory results (e.g. \cite{Bauer:2012pv}). Some
observables may depend strongly on the actual fit form used, e.g. the dipole
fits, being the least flexible, usually tend to give the smallest error.
However in \cite{A1:2013fsc} the dipole was found to potentially suffer from a large bias
if the true model is in fact not of dipole form. Most recent lattice studies
 quote results for the z-expansion. To be more specific the dipole or
 z-expansion \cite{Hill:2010yb} is given by
\begin{align}
	G_{\text{A}/\text{E}/\text{M}}^\text{dipole} (Q^2)&=
	\frac{a_{\text{A}/\text{E}/\text{M}}}{\Bigl( 1+ \frac{Q^2}{M^2}
	\Bigr)^2},\\
	G_{\text{A}/\text{E}/\text{M}}(Q^2) &= \sum \limits_{k=0}^\infty a_k\; z(Q^2)^k\, ,\\
        \intertext{with}
        z(Q^2) &=  \frac{\sqrt{t_\text{cut}+Q^2} - \sqrt{t_\text{cut}
	-t_0}}{\sqrt{t_\text{cut}+Q^2}+\sqrt{t_\text{cut} -t_0}},
\end{align}
where $t_\text{cut}$ is the branch cut for the respective form factor and $t_0$ is a
free parameter corresponding to the point in $t$ that maps onto $z=0$.
\begin{figure}[t]
        \begin{center}
        \includegraphics[width=.5\textwidth]{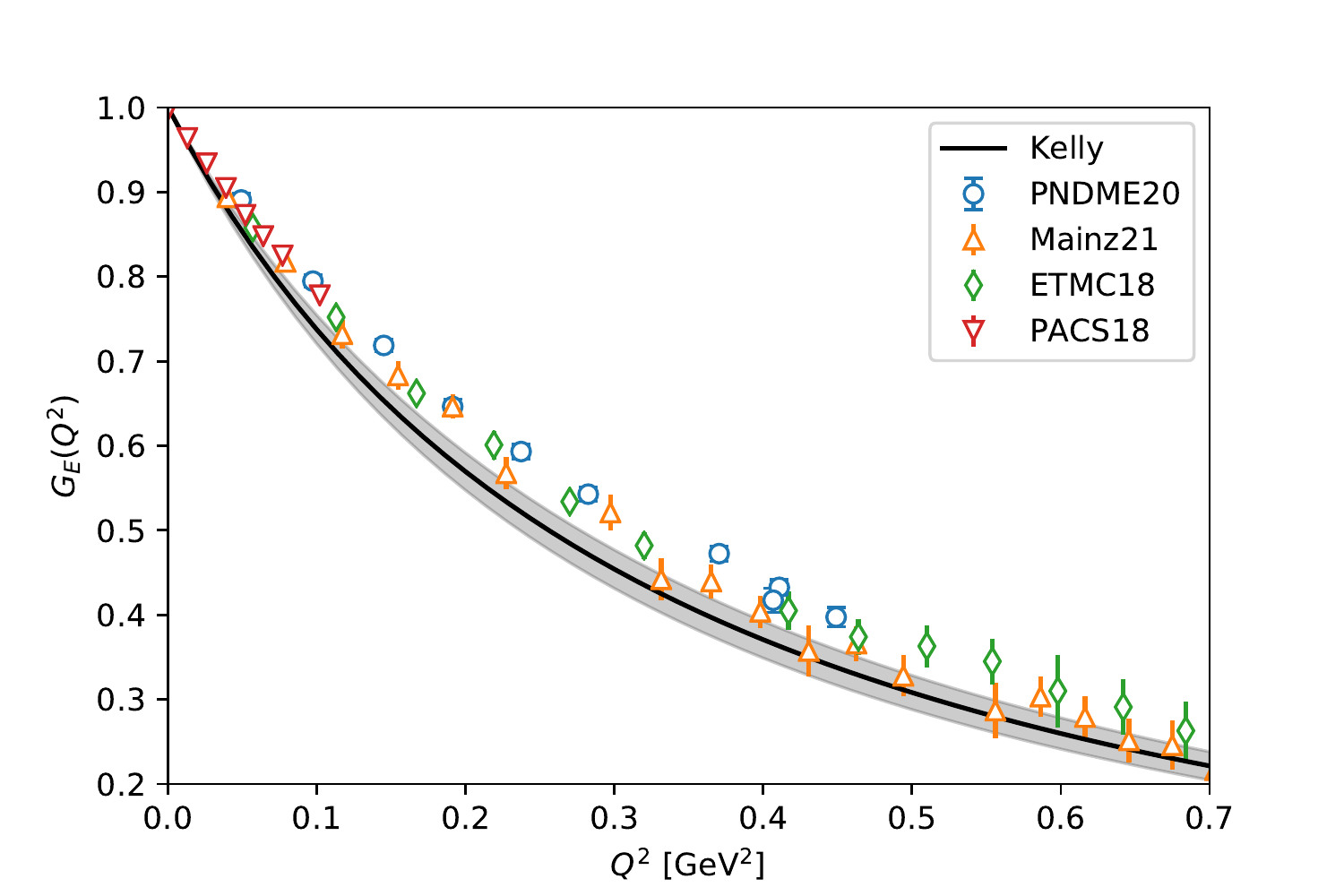}
\end{center}
\caption{Compilation of the isovector electric Sachs form factor for  physical pion mass ensembles, where
Kelly denotes the parametrization from \cite{Kelly:2004hm}, and the lattice data
are taken from Ref.~\cite{Jang:2019jkn} blue circles
(PNDME20), Ref.~\cite{Djukanovic:2021cgp} orange upwards triangles
(Mainz21), Ref.~\cite{Alexandrou:2018sjm} green diamonds (ETMC18) and Ref.~
\cite{Shintani:2018ozy} red
downward triangles (PACS18).}
\label{near_phys_pion_ge}
\end{figure}
The dipole is motivated
mostly by its simplicity and phenomenological success, whereas the $z$-expansion is
based on a conformal mapping, constructed such that one obtains the largest possible range of
convergence for the form factors, treated as a function of complex arguments. Here the cut
on the real axis, i.e. the threshold of (multi-)particle production, is accounted
for and amounts to  $\sqrt{t_{\text{cut}}}=3m_\pi (2m_\pi)$ in the axialvector (vector)
current case. For any parametrization interpolating between simulated points there is usually two ways the analysis
may proceed. In one approach the dipole or $z$-expansion fits are used to estimate derived
quantities, like radii, for fixed simulation parameters and subsequently
chiral and continuum fits are performed on
these. Alternatively the $z$-expansion formula may be amended with terms
parametrizing the chiral and discretization effects \cite{Jang:2019jkn}. In this
immediate form the number of data points that constrain the fit parameters is
larger compared to the two step process, and one may hope for more stable fits,
especially for the cases where one includes lattice spacing and finite volume
effects simultaneously. For $z$-expansion fits most studies obtain stable results
only to a low order in $z$ ($k\leq 2$),
stabilizing the fits with gaussian priors for the higher order
coefficients. The magnetic form factor is especially difficult as lattice
studies do not have a handle on the point at $Q^2=0$, leading to somewhat larger
uncertainties in the estimates of the magnetic moment and radius.
Another idea was pursued in Ref.~\cite{Djukanovic:2021cgp}, where formulae
based on an effective
field theory description are used \cite{Bauer:2012pv} to perform $Q^2$-, chiral- and
continuum extrapolations in one global analysis. In general the range of pion
masses for which the chiral extrapolations converge is not known, and cuts with
respect to the pion mass should be performed to assess potential bias. Most
recently there is also direct determination of the radius implementing the derivative on the
correlator level \cite{Alexandrou:2020aja,Ishikawa:2021eut} eliminating the modeling of the $Q^2$
dependence as a source of systematic uncertainty.

\begin{figure}[t]
	\begin{center}
	\includegraphics[width=.95\textwidth]{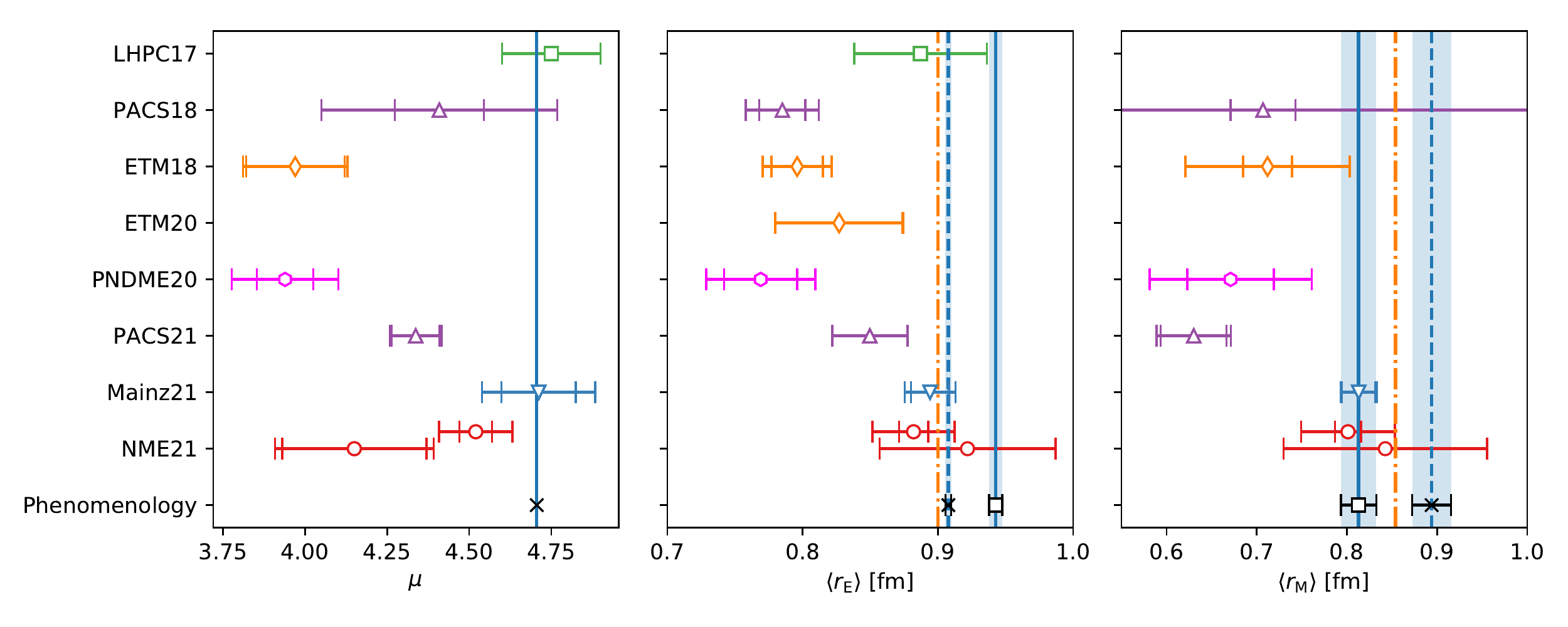}
\end{center}
\caption{Comparison of recent lattice determinations, from Ref.~\cite{Park:2021ypf} red circle
	(NME21), Ref.~\cite{Jang:2019jkn}  magenta hexagon (PNDME20),
	Refs.~\cite{Alexandrou:2018sjm,Alexandrou:2020aja} orange diamond
	(ETM18/20), Ref.~\cite{Djukanovic:2021cgp} blue downwards triangle
	(Mainz21), Refs.~\cite{Shintani:2018ozy,Ishikawa:2021eut} purple upwards triangle
	(PACS18/21), Ref.~\cite{Hasan:2017wwt} green squares (LHPC17), for the isovector 
magnetic moment, electric and magnetic radius, respectively. The yellow
dash-dotted lines correspond to the dispersive analysis of
Ref.~\cite{Lin:2021umz}.  The phenomenological values for the magnetic moment is
derived form the proton
and neutron values taken from PDG \cite{ParticleDataGroup:2020ssz}. The black
square denotes the  
electromagnetic radii corresponding to a value derived from Mainz/A1
\cite{A1:2010nsl} data
for the proton. Alternatively the crosses denote the derived values using, for $\langle r_\text{E}\rangle $ the CODATA2018 value of
the proton electric radius, and for the  $\langle r_\text{M}\rangle $ the
world data excluding Mainz/A1. In both cases the values for the neutron are
taken
from \cite{ParticleDataGroup:2020ssz}.} 
\label{fig_comp_isovec}
\end{figure}

The possible variations in the analysis can become quite numerous and it is not clear from the
start which variations really have a strong effect in the final observable. Therefor
choices have to be made in the course of the analysis to keep the number of
variations manageable. Once the number of different methods is fixed 
one still faces the problem of obtaining a final estimate from a range of
determinations. Most studies proceed to give a best estimate of one favored model with a
statistical error and an error based on variations of the analysis most
sensitive to one of the aforementioned sources of systematic uncertainty. While
for the statistical error resampling techniques are the de facto standard, for
the latter no unified approach exists and various methods are applied in the
literature. Some studies use the spread in the mean values of the variations
to assign a systematic error, e.g. the error is large enough to cover all mean values
of the variations. Another possibility is not to select a preferred model but to
perform model averages using information criteria like the Akaike Information
Criterion \cite{Akaike:IEEE:1100705}\footnote{In the case one model is clearly
	preferred by the AIC the model average effectively becomes a model
selection.  }. Even cuts performed on the data, e.g. in $Q^2$, $m_\pi L$ or
$m_\pi$, may be reinterpreted as a model selection problem
\cite{jay2020bayesian} and can be included in such an average. 

\begin{figure}[t]
	\begin{center}
	\includegraphics[height=.25\textheight]{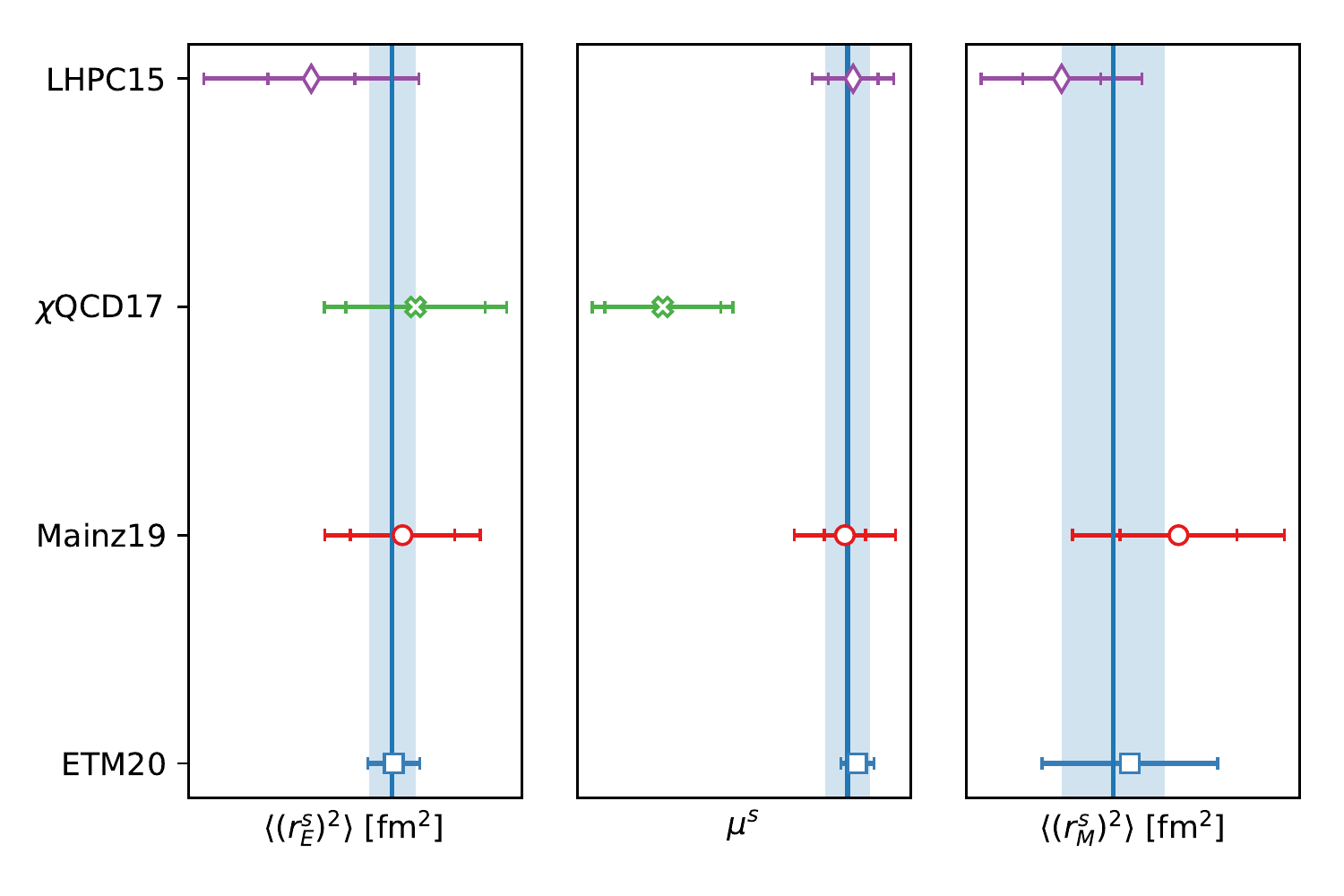}
\end{center}
\caption{Comparison of the strange electromagnetic form factor of the nucleon, data
	are from Ref.~\cite{Green:2015wqa} purple
	diamonds (LHPC15), Refs.~\cite{Sufian:2016pex,Sufian:2017osl} green crosses
	($\chi$QCD17), Ref.~\cite{Djukanovic:2019jtp} red circle (Mainz19), and
	Ref.~ \cite{Alexandrou:2019olr} blue squares (ETM20). The blue shaded
area is the PDG-style unconstrained weighted average.}
\label{fig_comp_strange}
\end{figure}

\section{Recent results}
In Fig.~\ref{near_phys_pion_ge} I show a compilation of the isovector electric
Sachs form factor for close to physical pion mass ensembles. For most of the $Q^2$
values the different extractions agree reasonably well within errors, where the
data from PNDME \cite{Jang:2019jkn} lies somewhat higher. The different
analysis treat excited states differently, e.g. for
Ref.~\cite{Djukanovic:2021cgp} data from the two-state fits with narrow  
priors, obtained from two-state fits, are shown. The effective form factors in
Ref.~\cite{Shintani:2018ozy}
have been extracted using plateau fits. In Ref.~\cite{Alexandrou:2018sjm} the ground state matrix
element was identified by two-state fits, however demanding consistency between determinations based on
two-state fits and the plateau method for several $t_\text{sep}$. For Ref.~
\cite{Jang:2019jkn} the excited state energy gaps from the two-point functions
were used as fixed input in the three-point function (data for a09m130W is shown). In
Fig.~\ref{fig_comp_isovec} I show a comparison of the latest results for the
isovector magnetic moment and electromagnetic radii of the nucleon. The
errors for the magnetic moment and radius are in general larger, mostly due to
the missing point at vanishing momentum transfer and the associated
extrapolation. It is also visible that
reproducing the magnetic moment already is a challenge for lattice
determinations. One may combine the experimentally available values for the
proton radius from $ep$-scattering of Ref.~\cite{A1:2010nsl} and from muonic hydrogen,
with the experimentally known values for the neutron to obtain a data driven
estimate of the isovector quantity, labelled \emph{Phenomenology} in
Fig.~\ref{fig_comp_isovec}, for both cases. For the electric radius we see that  most lattice
determination are comparatively low. While there is a tendency towards the smaller
value of the electric radius the $ep$-scattering value of Ref.~\cite{A1:2010nsl} cannot be ruled out. A
decisive statement about the proton radius puzzle is not yet possible. An
interesting observation is that
there seems to be a slight tension for the magnetic radius from the dispersive
analysis \cite{Lin:2021umz} and the lattice determination of
\cite{Djukanovic:2021cgp}, while there is good agreement for the electric radius
between the two.

\begin{figure}[t]
	\begin{center}
	\includegraphics[height=.25\textheight]{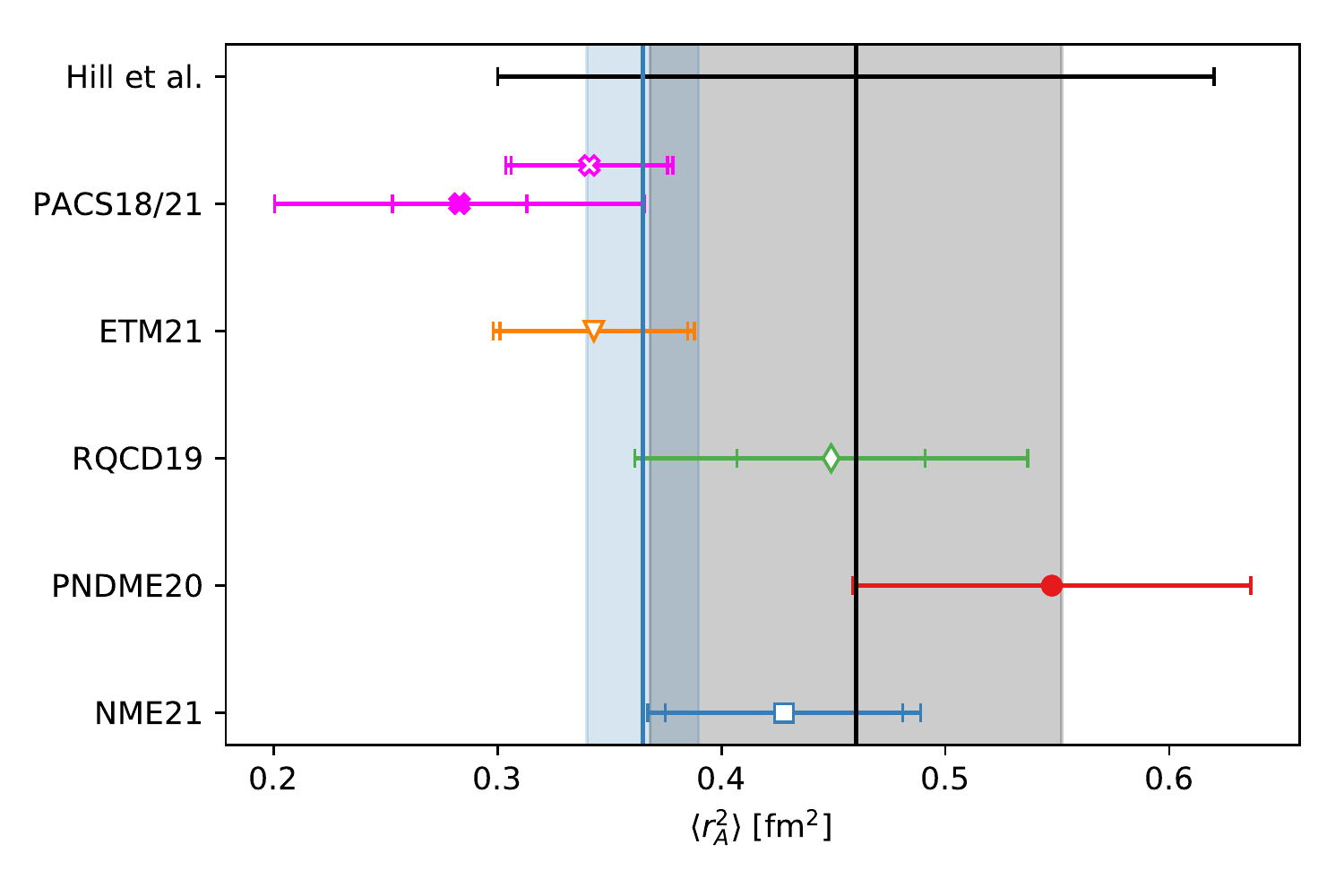}
	\caption{Comparison of lattice determinations of the axial radius
		$\langle r_\text{A}^2\rangle$, from Ref.~\cite{Park:2021ypf}
		blue square (NME21), Ref.~\cite{Jang:2019vkm} red circle 
		(PNDME20), Ref.~\cite{RQCD:2019jai} green diamond (RQCD19),
		Ref.~\cite{Alexandrou:2020okk} orange
		triangle (ETM21), Refs.~\cite{Shintani:2018ozy,Ishikawa:2021eut}
		magenta cross (PACS18/21). The point labeled Hill et al.
is  an average of the values obtained from z-expansion fits to neutrino
scattering and muon capture \cite{Hill:2017wgb}, with the grey band indicating a 20\% error for
$\langle r_\text{A}^2\rangle$. The blue band is  the unconstrained weighted average over the
data points with open symbols, i.e. the most recent more precise data sets which
quote a systematic error.}
\label{comp_axial_rad}
\end{center}
\end{figure}
For the case of the strange electromagnetic form factor I show a comparison in
Fig.~\ref{fig_comp_strange}. While for the strange magnetic moment there is some
tension between $\chi$QCD and the other estimates
\cite{Green:2015wqa,Djukanovic:2019jtp,Alexandrou:2019olr}, the electric and
magnetic radii agree very well amongst the different lattice determinations.
The blue shaded area in Fig.~\ref{fig_comp_strange} shows the PDG-style
unconstrained weighted average, where I added the systematic errors in
quadrature. This simple average reads
\begin{align}
	\mu^{s,\text{average}}&= -0.0193(54)\, , \\
	\langle (r_\text{E}^s)^2\rangle^{\text{average}} &=  -0.00484(54) \, \text{fm}^2,\\
	\langle (r_\text{M}^s)^2 \rangle^{\text{average}} &= -0.0167(53) \, \text{fm}^2.
	\label{weighted_average_strange}
\end{align}

A comparison  of recent results for the axial radius is shown in
Fig~\ref{comp_axial_rad}. All determinations individually meet the criterion of
20\% error for $r_\text{A}^2$, with respect to the statistical error, put forth to
render the axial radius uncertainty a subleading effect in the cross section of
quasi-elastic neutrino-nucleus scattering. The data depicted by open symbols agree within errors once the
systematic error is added in quadrature. The blue shaded area is the
unconstrained weighted average of these data points 
\cite{Park:2021ypf,RQCD:2019jai,Alexandrou:2020okk,Ishikawa:2021eut}, derived
from $z$-expansion or direct extractions, with systematic errors added in
quadrature, leading to a simple PDG-style average of 
\begin{align}
	\langle r_\text{A}^2\rangle^{\text{average}} = 0.365(25) \, \text{fm}^2\,.
	\label{ra_average}
\end{align}
Let me stress that this average does not include any quality criteria with
respect to the assessment of systematic errors. The systematics for the
individual lattice
determinations are  very different, e.g. in Ref.~\cite{RQCD:2019jai}
an extraction based on a dipole parametrization of the $Q^2$ behavior gives a
considerably smaller value, whereas in \cite{Ishikawa:2021eut} the radius is obtained
implementing the derivative directly. 

\section{Summary}
Excited-state contaminations
remain the most dominant source of systematic uncertainty for the form factors
of the nucleon. Most analysis use multi-state fits to account for the effect of excited states
at the correlator level, with a varying degree of prior knowledge applied to
guide and stabilize procedures. Alternatively the summation method is used,
where a comparable level of precision is usually achieved only with the extension
of this method to include smaller source-sink separations below 1 fm. In the absence of
a superior method for dealing with excited states, or more detailed knowledge of
the excitation spectrum of the nucleon, it is important to show for
every analysis that residual effects are under control. To that end one ideally
performs several variations designed to elucidate the influence of the various
systematics. Averaging the results of these variations based on AIC
weights could provide an efficient way to establish the effects of systematics on the
final observable. 

In recent studies the effect of excited states was observed to be amplified in
the case of the axialvector current. While the extractions for the axial
radius seem to have reached the level of accuracy needed for example for the
analysis of neutrino-nucleus scattering, further corroboration of the results is
desirable to increase confidence that indeed all systematics are well under control.

The strange electromagnetic form factors of the nucleon, a key ingredient in the analysis
of low energy parity-violation experiment, are in good agreement between the
different lattice determinations, resulting in non-zero values for the strange
magnetic moment and electromagnetic radii. More recently in
Ref.~\cite{Alexandrou:2021wzv} the strange axial form factor has been
determined at physical pion mass. The systematic
uncertainties in this quantity need to be further investigated.

Even though there is strong indication for a smaller value of the electric
radius of the proton, as determined from muonic hydrogen, lattice determinations
at the current level of statistics and systematics may not rule out the
$ep$-scattering value of \cite{A1:2010nsl} with a high level of confidence. For
an improved extraction, especially  of the magnetic radius, more data points at
low $Q^2$ are needed. 

\bibliographystyle{JHEP}
\bibliography{skeleton}

\end{document}